\pdfoutput=1
\documentclass[doublecol]{epl2} 

\usepackage{graphicx}
\usepackage{amssymb}
\def\artitle#1{} 
\long\def\omitt#1{}
\usepackage{TomspeakEPL} 
\newcommand{\be}{\begin{equation}}
\newcommand{\ble}[1]{\begin{equation}\label{#1}}
\newcommand{\ee}{\end{equation}}

\title{Robust fadeout profile of an evaporation stain}
\author{ T. A. Witten} 
\institute{James Franck Institute, University of Chicago, Chicago IL 60637 USA.}


\abstract{
We propose an explanation for the commonly-seen fading in the density of a stain remaining after a droplet has dried on a surface.  The density decreases as a power $p$ of the distance from the edge.  For thin, dilute drops of general shape 
this power is determined by a flow stagnation point in the distant 
interior of the drop.  
The power $p$ depends on the local evaporation rate $J(0)$ at the stagnation point and the liquid depth $h(0)$ there: $p = 1 - 2~ (h(0)/\bar h)(\bar J/J(0))$, where $\bar h$ and $\bar J$ are averages over the drop surface. 
}

\begin{document}

\maketitle
\section{Introduction} \label{sec:introduction}

\begin{figure}
\onefigure[width=\hsize]{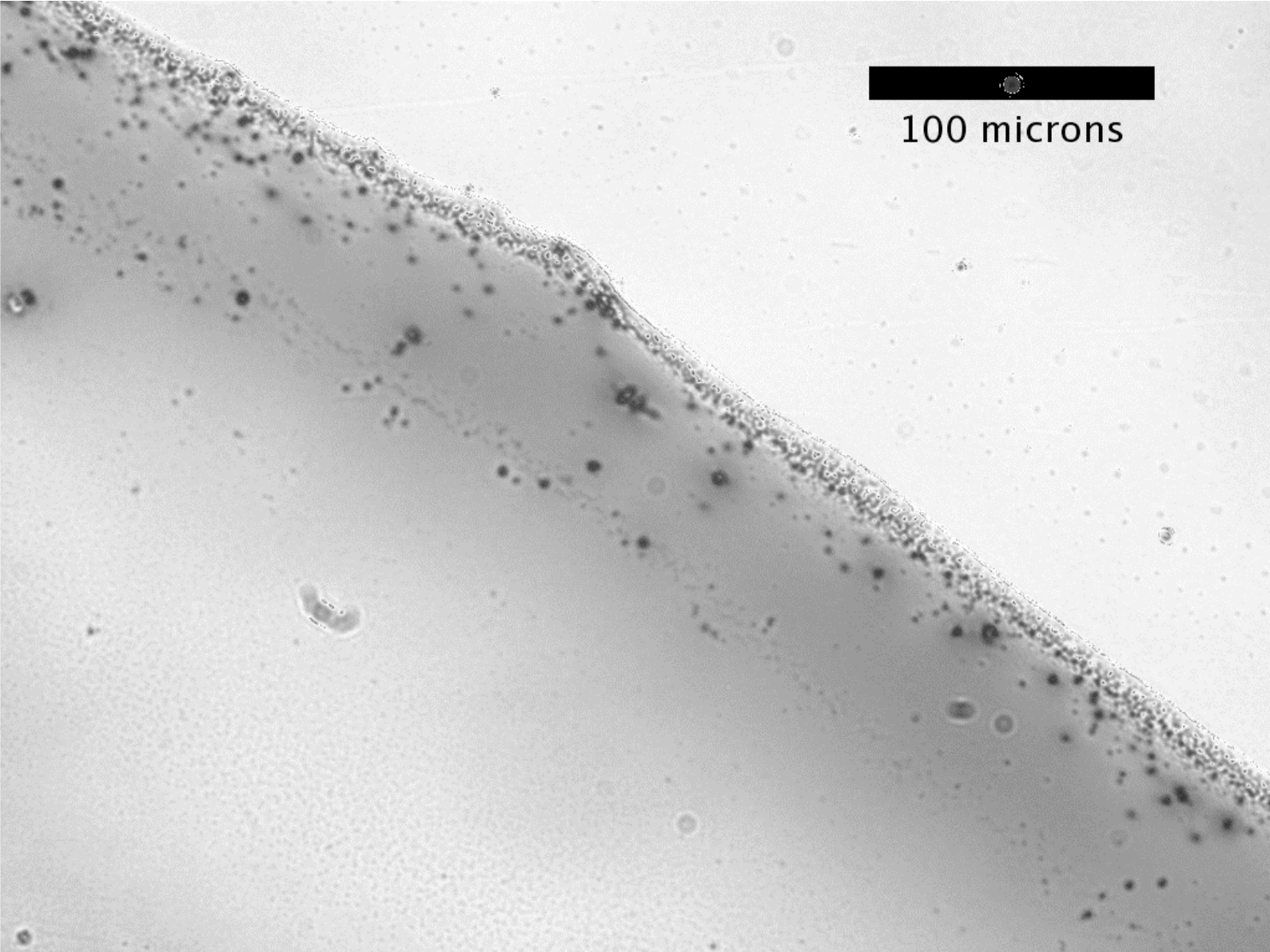}

\caption{Optical micrograph of the edge of a 5 mm drop of black ink diluted with water, deposited on a glass microscope slide and dried in air. Continuous decrease of image density occurs as one moves downward from the edge of the drop towards the center.  The discrete particles and other patterns are not addressed here.}
\label{fig:concentrated}
\end{figure}

			Recent years have witnessed startling 
new mechanisms for creating predictable, self-organized structures from long-known classical principles\cite{structures}.  Among these newly recognized mechanisms is the deposition of solute in an evaporating drop of liquid\cite{Deegan.Nature}.  Evaporation entails a specific flow pattern of the liquid which in turn concentrates the solute onto the perimeter of the drop.  Many universal features of the resulting deposition have been tested\cite{DeeganPRE, KijiyaDoi2008} and exploited for applications\cite{Larson, Kimura, Smalyukh,  J.Conrad.stratification, BertelootLimat2008}.  However, several prevalent features of the density profile of the deposition still await explanation\cite{Deegan.thesis}.  Here we address one such feature: the decrease or fadeout of density with distance from the perimeter, as shown in Fig. \ref{fig:concentrated}.  Our mechanism predicts a power law fadeout whose exponent depends on two controllable features of the evaporation profile and droplet shape.

			 The drop of Fig. 1 contained 
 a volatile solvent that partially wet the solid substrate; this solvent leaves the drop over time via evaporation.  \revision{The drop \omitt{5/26} also} contained nonvolatile molecular or colloidal solutes.  These may be carried along by any lateral flow of the solvent as it dries, but they are not carried away when the solvent evaporates. 
Such lateral flow occurs because a) the perimeter or contact line is typically pinned by irregularities in the surface and previous solute deposition, b) the free surface of the liquid takes the equilibrium shape of constant mean curvature dictated by its surface tension, and c) the local change of volume dictated by the thinning 
of this equilibrium shape is not matched by the local loss due to evaporation.  To supply the volume needed for evaporation, lateral flow is required.  Near the perimeter the evaporative loss greatly outweighs the supply due to local thinning; thus, the lateral flow is strong.  \revision{It is sufficiently strong}\omitt{5/26} to carry any point of the drop's interior to the perimeter during the drying time\cite{Deegan.Nature}.  For a thin, circular drop with evaporation controlled by air diffusion, the flow field may be readily determined and the consequent accumulation of solute with time deduced\cite{DeeganPRE}: in this base case the mass deposited at the perimeter in time $t$ varies initially 
as the $t^{4/3}$.  

			While these facts explain the strong accumulation of the solute at the perimeter, they do not explain the profile of solute density seen in evaporated drops.  From a well-defined outer edge, this density rises steeply as one proceeds inward, reaches a maximum, and then fades gradually away.  The width \revision{$w$} of the deposit increases with the initial concentration of solute.  To understand this profile one must relate the solute's motion to its concentration.    In 1996 Dupont\cite{DupontPrivate} proposed a simple mechanism to account for the effect of concentration.  Recognizing that solute motion must be inhibited as its concentration increases, he postulated that the solute simply stops moving when its local volume fraction $\phi$ exceeds a threshold value $\phi_c$, while solute at lower volume fraction is carried passively.  Following earlier work by Popov\cite{Popov.drying}, Zheng has recently analyzed the surface\omitt{new 5/25} density profile resulting from the Dupont mechanism in two important cases\cite{RuiZheng}. \revision{Zheng defines the surface density $\Sigma$ as the volume of solvent per unit surface area measured after the solvent has evaporated\omitt{new 5/25}.} For our base case this $\Sigma(x)$ varies with distance $x$ from the perimeter as $x^{-7}$.  

			Below we simplify and generalize Zheng's analysis to a broad class of evaporation profiles and droplet shapes. We first define and justify the simplified physical system to be analyzed.  We recall\cite{Popov.drying, RuiZheng} that the density profile is controlled by the advance of the solidification front separating the mobile from the immobile solute.  We observe that the fadeout deposition corresponds to a late stage of evaporation controlled by a stagnation region of the flow near the middle of the drop.  We then 
show that the fadeout exponent $p$ is entirely determined by a dimensionless ``Poisson ratio" characterizing the distant stagnation point. 
We first consider circular droplets and then more general shapes.  

			A \revision{sessile drop wets a circular region} of fixed radius $X$ contains a solute at initial volume fraction $\phi_i \muchlessthan \phi_c$, uniformly dispersed in the fluid.  \revision{Since the solute is dilute, the width $w$ of the deposit will be arbitrarily narrow in comparison to the drop radius $X$.}  \omitt{new 5/26/}  The initial contact angle is small, so that the height $h(r)$ at distance $r$ from the center is much smaller than $X$ everywhere.  We take $X$ itself to be small enough that gravitational distortions of the shape are negligible.  Thus the initial $h(r)$ can be written \revision{$h_i(r) = H_i~(1-(r/X)^2)$}\omitt{OK \quad}, where $H$ is the center height.  At time $t = 0$ evaporation begins.  The volume removed per unit time and per unit area from the surface at $r$ is denoted $J(r)$ (Fig. 2 inset).  This $J$ depends on the thermodynamics of evaporation and the environment of the drop, but not on time\cite{DeeganPRE}.  By increasing the saturation of the surrounding atmosphere, we may make $J$ as small as we wish.  We thus choose $J$ to be so small that kinetic effects in the liquid such as viscous dissipation are negligible: the flow is quasistatic. 
We choose solute particles that do not absorb on the surfaces.  Their diffusion during the drying time $T$ is \revision{made} \omitt{5/26} negligible on the scale of $X$ but substantial on the scale of $H$, so that the solute remains uniformly dispersed in depth during the evaporation.   Further, the solute particles must be sufficiently small or polydisperse that the solute concentration may be treated as a continuum and discrete-particle layering effects\cite{Stone.layering} are avoided.  For the moment, 
we suppose that the solute is \revision{perfectly} \omitt{5/25 } {\it compressible}, so that the immobilized solute does not alter the shape of the fluid surface.  

			As the evaporation proceeds, the central height $H$ decreases to zero linearly in time (since the rate of volume loss is constant).  A lateral flow at the depth-averaged speed $v(r, t)$ arises in order to restore the imbalance of volume loss between $J(r)$ and $\dot h(r, t)$: \revision{$\del \cdot (v~ h) = -\dot h - J$}\omitt{Bk130P83f fixed \quad}. Since $J(r)$ and \revision{$\dot h(r) = -h_i(r)/T$}\omitt{Bk130P83f fixed\quad} on the right remain fixed in time, the $v~h $ on the left side is time-independent as well.  However, the solute profile $\phi(r)$ undergoes qualitative changes from its initial uniform state as drying proceeds.  This evolution gives rise to changing deposition over time.  The two limiting regimes of interest are the {\it early regime} $t/T \muchlessthan 1$ and the {\it late regime} $(1-t/T)\muchlessthan 1$.  \revision{In this late time regime \omitt{5/25} the fraction of remaining solvent is small and solute from the central region will have reached the perimeter.  On the other hand, the remaining solvent may still far exceed the amount of solute.}  We shall suppose that the initial volume fraction $\phi_i$ is so small that the average volume fraction remains much smaller than $\phi_c$ during the late as well as the early regimes.  We shall not consider the {\it final regime}, in which $\phi$ has become comparable to $\phi_c$, since it contributes arbitrarily little to the deposition profile of interest.  \revision{Also, by avoiding the final time regime we avoid complications like concave meniscus, contact line depinning and viscous stresses.} \omitt{5/26}
				\omitt{must solute be compressible? See NOTE 2 after \\end }

\section{Deposition front} \label{sec:deposition}

	If the solute were transported purely passively, a nonzero mass of solute would be transported to the perimeter over time.  Thus the density there would immediately become infinite.  Imposing the Dupont constraint $\phi< \phi_c$ prevents this infinite density.  Immediately a zone of immobilized solute appears at the perimeter and begins to expand inward.  Outside this immobilization front, lateral motion ceases and the surface density $\Sigma(x)$ becomes fixed in time.  However, the fixed solute 
does not alter the outward \revision{flow}\omitt{5/26} profile \revision{$v~h$}, which depends only on the height and evaporation profiles.  Under the quasistatic conditions we have chosen 
this $v ~h$ 
remains unaltered by the solute.  			
\begin{figure}
\onefigure[width=\hsize]{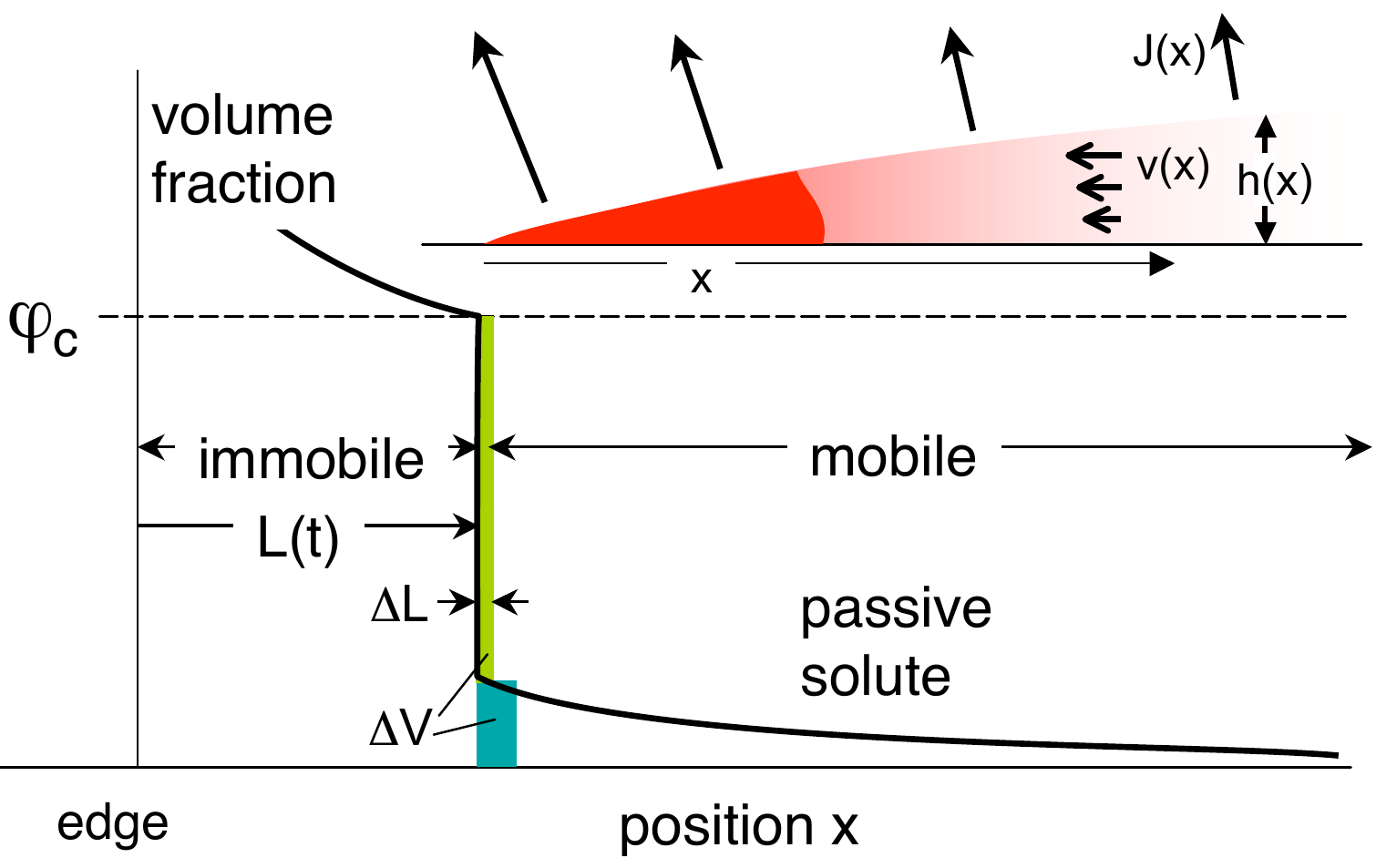}
\caption{(color online) Inset: vertical section near the edge of a drop.  Immobile region (solid color) evaporative current $J$ lateral velocity $v$ and height $h$ are shown.  Main figure: sketch of volume fraction $\phi$ vs distance from perimeter $x$ in the Dupont model\cite{RuiZheng}.  To the right, the volume fraction is unaffected by immobilization.  \revision{To the left}, above the volume fraction $\phi_c$ indicated by a horizontal line, the solute is immobile. \revision{Here the volume fraction $\phi(x)$is larger than $\phi_c$ only because the height at $x$ has decreased after the deposition occurred there.} The immobilized region is bounded by the immobilization front at $L(t)$.  
The solute volume \revision{$\Delta V$}\omitt{5/25}  advected to the front 
in an interval $\Delta t$ is shown by the lower dark-colored bar.  This increment of solute is sufficient to increase the concentration just ahead of the front to $\phi_c$ for a width $\Delta L$ indicated by upper light-colored bar.  Accordingly, the immobilization front advances by $\Delta L$.  
For the dilute case treated in the text, mobile volume fraction is very small compared to $\phi_c$.}
\label{fig:ShockFront}
\end{figure}

The motion of the front can be found by a mass-balance argument, as sketched in Fig. 2.  Denoting its current distance from the perimeter by $L(t)$, its advance $\Delta L$ in a time $\Delta t$ is proportional to the amount of solute carried to the front from the interior during $\Delta t$.  The solute reaching the perimeter at time $t$ originated from some radius $r_i(t)$ in the interior of the drop.  In time $\Delta t$, material is advected to $r_i$ from a point $r_i - v(r_i, t) \Delta t$ just inside $r_i$.  
\omitt{this new material much thus reach the boundary at time $t + \Delta t$.}
	Thus\cite{DeeganPRE, RuiZheng} \revision{$\dot r_i = -v(r_i, t)$}\omitt{Bk130P83f \quad}.  
The volume $\Delta V$ of solute reaching the perimeter at time $\Delta t$ is thus $\phi_i ~2\pi r_i ~h_i(r_i)~ \Delta r_i$\omitt{Bk130P83f \quad} or $\phi_i~ 2\pi r_i h_i(r_i) v(r_i, t) \Delta t$\omitt{Bk130P83f \quad}.  The volume reaching the immobilization front, just inside the perimeter, is the same in the limit $L \muchlessthan X$.  We determine the motion of $L$ by noting that the volume fraction in the $\Delta L$ increment must be $\phi_c$: 
\revision{\ble{eqn:label}
\phi_c ~ \Delta L ~ h(X - L, t) ~2\pi X = \Delta V 
\ee
}
\omitt{Bk130P83f \quad} \omitt{argument of h fixed 5/26}
Since \revision{$h(X-L, t) \goesto 2 L~ H/X$} for $L \muchlessthan X$\omitt{Bk130P83f \quad} \omitt{fixed argument of h 5/26}, this condition yields
\ble{eqn:front}
L~\dot L = {r_i h_i(r_i)\over H(t)}~ {\phi_i \over \phi_c}~ \half v(r_i, t) 
\ee
\omitt{Bk130P83f \quad}
The quantities on the right side are known.  This equation thus determines the motion of the front. Given $L(t)$ we may determine the surface density $\Sigma(x)$.  This density is fixed at the moment $t_L(x)$ when the front crosses $x$.  At that moment \revision{$\Sigma(L(t)) = \phi_c~h(X-L, t)$.}  
\omitt{Bk130P83f \quad}\omitt{fixed argument of h 5/26}

\section{Stagnation flow} \label{sec:stagnation}
We now consider how $\Sigma(x)$ is determined in the late-time limit.
In this limit the initial radius $r_i(t)$ is close to the center where $r=0$.  Here the fluid's \revision{vector}\omitt{5/25} velocity $v(r) \proportionalto r$\omitt{Bk130P83f \quad}.  We may determine $v(r)$ by considering the evolution of a small concentric cylinder of liquid of radius $r$ and area $A$. 
The expansion $\Delta A$ in this area is driven by the local evaporation current
$J(0)$ and height $h(0)$ (Fig. 3, left inset). 
 In time $\Delta t$, the cylinder loses volume $\Delta \Omega_J = J(0) ~A \Delta t$\omitt{Bk130P83f \quad} by evaporation.  It loses volume through the decrease in height by an amount \revision{$\Delta \Omega_h = -A  \Delta h$}\omitt{Bk130P83f \quad}.  The loss due to evaporation is not sufficient to reduce the height by the required amount.  Thus the cylinder's area must expand by an amount $\Delta A$ such that \revision{$h(0) ~\Delta A = \Delta \Omega_h - \Delta \Omega_J = A~ (-\Delta h(0)) (1 + J(0)/\dot h(0))$}\omitt{Bk130P83f \quad}.  
 Evidently $\dot h(0)$ and $J(0)$ are related, since $J$ is what causes 
$h$ to change.  This relationship can be expressed in terms of the area average of $J$, denoted by $\bar J$, and the corresponding height average $\bar h$: $\bar J = -\dot{\bar h} = \bar h_i/T$\omitt{Bk130P83f \quad}.  Also $\dot h(0)$ is given by \revision{$- h_i(0)/T$}\omitt{Bk130P83f \quad} so that \revision{$J(0)/\dot h(0)  = -(J(0)/\bar J)~ (\bar h_i / h_i(0))$}\omitt{Bk130P83f \quad}. Noting that  $\bar h_i / h_i(0)=\bar h / h(0)$\omitt{Bk130P83f \quad}, we infer 
\ble{eqn:poisson}
\Delta A /A =  [1  - \alpha]~ ~ (-\Delta h / h), 
\ee	
where $\alpha \definedas (J(0)/\bar J) ~(\bar h / h(0))$. \omitt{Bk130P83f \quad}
The relative gain of area is proportional to the relative loss of height, as in the compression of an \revision{elastic solid\cite{solid.mechanics.text} with Poisson ratio $\nu = [1-\alpha]/2$}. \omitt{5/26} For typical evaporating drops $J(0)$ is positive, so that $2\nu < 1$.  In order to have an outward evaporating flow with a positive $\Delta A$, $\nu$ must be greater than zero.

	This ``Poisson ratio" $\nu$ describing the stagnation flow controls the motion of the immobilization front and thence the deposit profile $\Sigma(x)$.  Evidently $\dot A = -2\nu A~ \dot h/h$.\omitt{Bk130P83f \quad}
The $-h/\dot h$ is simply the remaining time $T-t \definedas u$.
From $\dot A$ we determine $\dot r_i$ via $\dot A_i/A_i = 2 v_i(r_i)/r_i  = -2 \dot r_i/r_i$\omitt{Bk130P83f \quad}.  Thus Eq. \ref{eqn:poisson} becomes $-2 \dot r_i/r_i = 2\nu /u$\omitt{Bk130P83f \quad}, and $r_i \proportionalto u^{\nu}$\omitt{Bk130P83f \quad}.  We may now evaluate the front equation (\ref{eqn:front}) noting that $h_i(r_i)\goesto H_i$\omitt{Bk130P83f \quad}, $H(t) = (u/T)~ H_i$ and \revision{$v(r_i, t) = d r_i/du \proportionalto u^{\nu - 1}$}\omitt{Bk130P83f \quad}.   
\ble{eqn:front2}
L~\dot L \proportionalto ~  u^{2\nu - 2} ,
\ee
\omitt{Bk130P83f \quad}
so that $L \proportionalto u^{\nu - 1/2}$\omitt{Bk130P83f \quad}.  Since $\nu < \half$ as noted above,  $L$ diverges as a fractional power of the remaining time.  \revision{This means that the front extends inward to distances much greater than the small deposit width $w$; our theory ignores the final regime where $L$ grows comparable to the radius $X$}.\omitt{5/26}
\omitt{
\ble{eqn:label}d L^2/du \goesas u^{2\nu - 2} 
	L^2 \goesas u^{2\nu - 1}
	L \goesas u^{\nu - 1/2}
\ee}

	From the front's motion the fadeout profile follows directly.  As noted above $\Sigma(x) = \phi_c h(x, t_L(x))$\omitt{Bk130P83f \quad}, so that $\Sigma(x) \goesas x^{(\nu + \half)/(\nu - \half)}$\omitt{Bk130P83f \quad}.  This amounts to $\Sigma(x) \goesas x^{1-2/\alpha}$\omitt{Bk130P83f \quad}.  For the base case described above, $h(0) = 2 \bar h$\omitt{Bk130P83f \quad} and $J(0) = \half \bar J$\omitt{Bk130P83f \quad}\cite{DeeganPRE}, so that $\alpha = 1/4$\omitt{Bk130P83f \quad} and $\Sigma \goesas x^{-7}$\omitt{Bk130P83f \quad}, in agreement with the detailed calculation of Ref. \cite{RuiZheng}.  For uniform evaporation with $J(0) = \bar J$\omitt{Bk130P83f \quad}, we find $\alpha = 1/2$\omitt{Bk130P83f \quad} and $\Sigma \goesas x^{-3}$\omitt{Bk130P83f \quad}, again in agreement with Ref. \cite{RuiZheng}. Similarly, for a 
gravity-flattened drop with $h(0) = \bar h$\omitt{Bk130P83f \quad} with $J(0) = \half \bar J$\omitt{Bk130P83f \quad}, we find $\alpha=1/2$\omitt{Bk130P83f \quad} so that again $\Sigma \goesas x^{-3}$\omitt{OK \quad}.  Finally, a flattened drop with uniform evaporation yields nominally $\alpha = 1$\omitt{Bk130P83f \quad} and $\Sigma \goesas x^{-1}$\omitt{Bk130P83f \quad}.  However this case lies marginally beyond the scope of our assumptions, since there is no reason here for a global outward flow of solvent.  
\omitt{
\ble{eqn:label}
\Sigma(x) \goesas h(x, t_L(x)) \goesas x u \goesas x ~ x^{1/(\nu - \half)}
\ee\ble{eqn:label}
\Sigma(x) \goesas x^{(\nu - \half)/(\nu - \half)}~ x^{1/(\nu - \half)}
= x^{(\nu + \half)/(\nu - \half)}
\ee
\omitt{check \quad}
$2\nu = 1 - \alpha$\omitt{check \quad}
\ble{eqn:label}
(\nu + \half)/(\nu - \half) = (\alpha-2)/\alpha = (1-2/\alpha)
\ee
}
\section{General shapes} \label{sec:general}

	The reasoning above may readily be generalized to non-circular drops. As before, the evaporative current $J(\vector r)$ is constant in time, and the height profile retains its initial shape during evaporation so that $h(\vector r, t) = (u/T) h_i(\vector r)$\omitt{ok \quad}.  As before, each point at position $y$ on the perimeter accumulates solute and has an immobilization front at $L(y, t)$.  
We denote the height at the front as $h(y, L, t)$.\omitt{referee complained \quad} 
The front advances because of solute advected from the interior.  As before each point on the perimeter receives solute at time $t$ from a point initially at $\vector r_i(y, t)$ in the interior.  Again this interior point moves inward in time according to  \revision{$\dot r_i(y, t) = -\vector v(\vector r_i, t)$} \omitt{Bk130P83f \quad}.  The velocity is driven by a scalar pressure field which decreases along streamlines. 
	
\begin{figure}
\onefigure[width=\hsize]{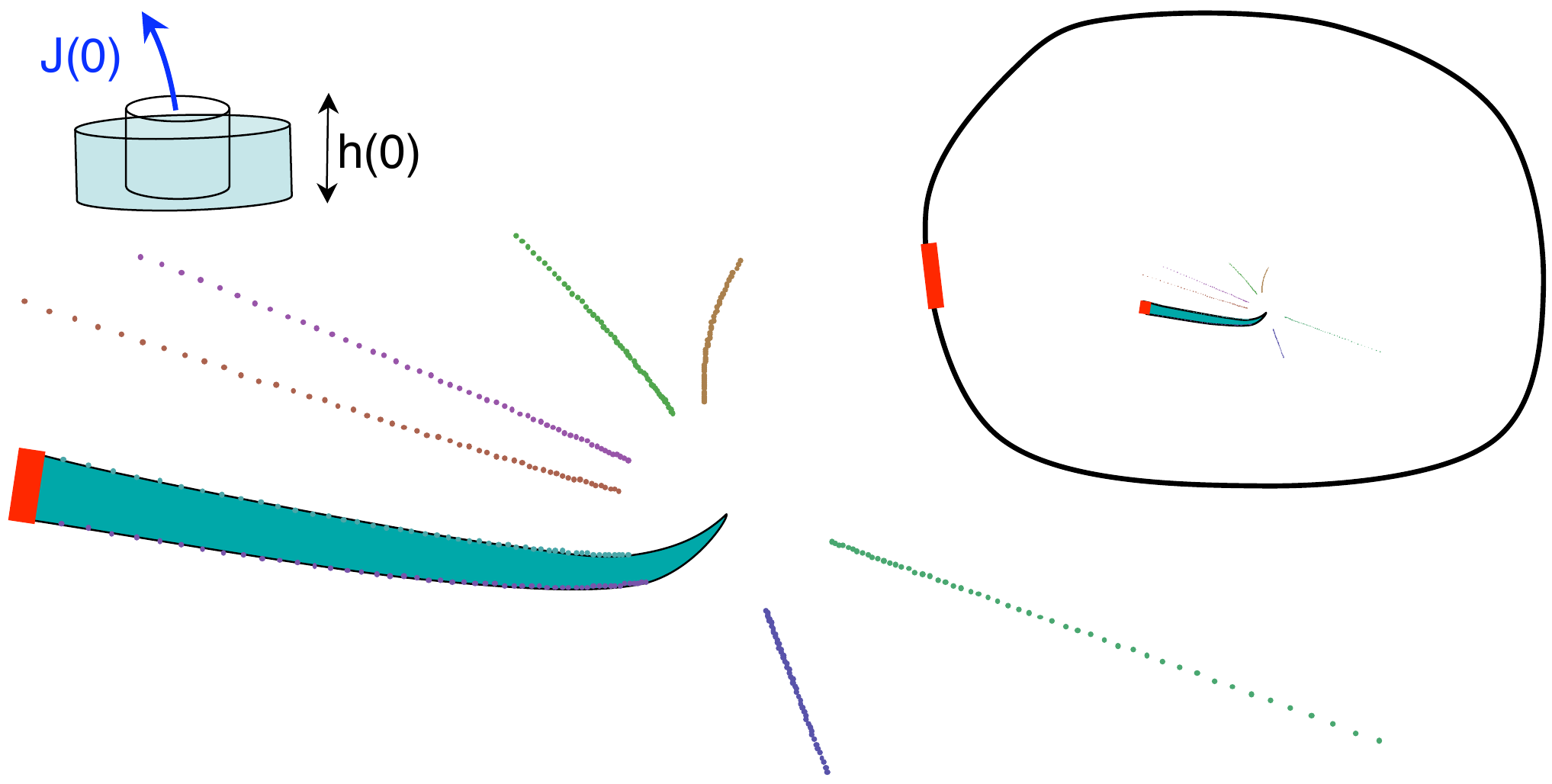}
\caption{(color online) Sketch of the stagnation flow in an evaporating drop of generic shape shown in the right inset.  A segment $\Delta y$ at the boundary is marked in color; its initial pre-image at time $t$ earlier is shown by the colored segment at the left of the main figure.  The wedge-shaped area traced out by this segment is shown as a shaded region. Selected streamlines from a small circle around the stagnation point are shown as rows of dots.  The two straight lines at lower right show the principal flow directions.  Opposite to each straight line is a pair of lines, showing the convergence of streamlines in the maximum-flow direction and divergence in the minimum-flow direction.  Left inset shows the advance of a small cylinder of fluid (unshaded) to the shaded region at the stagnation point.  The evaporation current $J(0)$ and height $h(0)$ are indicated.}
\label{fig:stagnation}
	\end{figure}

	The $r_i(t)$ extends inward with time and eventually reaches a stagnation point of highest pressure.  As with the circular drop, this stagnation region is the origin of the material that creates the fadeout profile.  Here the stagnation flow is anisotropic as shown in Fig. \ref{fig:stagnation}.
	
	To determine the fadeout profile, we consider the pre-image of a segment $\Delta y$ at time $t$, denoted $\Delta y_i(t)$.  We may determine its motion by considering a wedge-shaped region of the drop traced out by $\Delta y_i(t)$ as $t \goesto T$\omitt{Bk130P83f \quad} (Fig. \ref{fig:stagnation}).  We then analyze the increase 
in the area $\Delta A$ of this wedge with time.  This area changes for the same reasons as considered above, so that it too obeys $\Delta A / A = [1 - \alpha] ~(-\Delta h(0)/h(0))$\omitt{Bk130P83f \quad} with $\alpha \definedas (J(0)/\bar J) (\bar h/h(0))$. The solute in the area $\Delta A$ is evidently carried to the boundary segment $\Delta y$ in time $\Delta t$.  As before, this newly added solute creates an immobile region of volume fraction $\phi_c$ and width $\Delta L$ given by $\phi_c \Delta y~ \Delta L ~h(y, L, t) 
= \phi_i h_i(0) \Delta A$\omitt{Bk130P83f \quad}.  Since $\Delta A$ is given by the same expression as in the circular case, the growth of $L$ with time is also the same: $L(t)\goesas u^{\nu - \half}$ ({\it cf.} Eq. \ref{eqn:front2})\omitt{OK \quad}. Finally $\Sigma(x)= \phi_c h(x, y, t) \goesto x^{1 -2/\alpha}$\omitt{OK \quad}, where $\alpha = (J(0)/\bar J) (\bar h/h(0))$\omitt{OK \quad}.
	
\section{Discussion} \label{sec:discussion}
			
			We infer that the fadeout-profile of an evaporating drop is robust and general.  Of the many known forms of singular pattern formation, the fadeout mechanism described here is notable for its remotely controlled quality.  The flow properties at a single point of the drop control the power-law profile of deposition in the remote region at the perimeter.  Under the conditions we have assumed, the fadeout follows the same power law over the whole perimeter (though the amplitude depends on position).  Though the flow leading to the fadeout is generally difficult to calculate, the predicted power depends only on two simple features of the system: the normalized evaporation current and normalized height at the stagnation point.  These may be deliberately controlled. For example, the evaporation profile $J(r)$ can be controlled by putting a perforated lid above the drop\cite{J.Conrad.stratification}.  Thus one may shape materials on small scales where explicit molding or machining are not feasible,  
\omitt{new 3/26:}
Another aspect of capillary shaping of solutes was recently demonstrated by Vakarelski \etal\cite{Vakarelski}. The fadeout mechanism described here implements a form of shaping that complements Ref. \cite{Vakarelski}, 
for use when smooth and controllable gradients are desired.  
						
			Naturally the accuracy of these asymptotic predictions is limited in practice.  If one ventures too far from the circular 
shapes and $J$ profiles treated by Zheng, the asymptotic regime likely shrinks.  Ultimately the qualitative situation of a smooth, outward flow from a single stagnation point would be expected to break down.  Moreover, in order to find the controlling ratios $J(0)/\bar J$ and $h(0)/\bar h$ one must know the position of the stagnation point. For general geometries, this point is not known {\it a priori}. 
			
			The abrupt Dupont rule for immobilization appears unrealistic at first sight.  In reality, the loss of mobility with increasing volume fraction is surely more gradual, occurring over perhaps 
an order of magnitude of volume fraction leading to $\phi_c$.  However, in the dilute limit analyzed here, this abruptness has little effect.  Over the time range of interest, the volume fraction throughout the mobile region is indefinitely smaller than $\phi_c$; thus 
the concentration range in question occurs over an arbitrarily small spatial distance.  Therefore 
a smoother onset of immobility would not have affected the deposition.  
			
			Apart from these geometric effects, one expects material effects to alter the deposition.  Brownian motion of the solute\cite{Larson}, adsorption on the surfaces, stratification of the solute\cite{J.Conrad.stratification}  and depinning of the perimeter\cite{DeeganPRE} must distort the deposition to some degree.  The final regime of evaporation, where the mobile volume fraction approaches $\phi_c$, must also cause distortion.  \revision{In many experimental conditions these competing effects can easily dominate the ones considered here\cite{Larson}. Yet the competing effects can readily be reduced in order to isolate the essential effects treated here.}\omitt{5/25}  
						
			Our supposition of a \revision{compressible} \omitt{5/25} solute requires special discussion.  We assumed that the solute is perfectly compressible 
even in the immobilized state.  Typical polymeric solutes come close to this ideal, since they interfere with each other's motion even when their volume fraction is only a percent or less\cite{Polymer.Viscosity}. \revision{The polymers can thus resist the advective flow $v$ while still compressing vertically to many times their immobilized concentration.}  \omitt{5/25} However, many solutes are not compressible.  These would build up a deposit whose height could not change with time, thus forming a dam around the enclosed fluid.  The dam perturbs the height profile of the mobile fluid within, contradicting our assumption\cite{Popov.drying}.  In principle, this altered height profile creates an altered flow profile and this complicates the analysis.  \revision{It must have a strong effect in the final time regime, where the fluid surface would eventually become concave.} \omitt{5/25}  If we avoid the final time regime, then the height is only appreciably altered close to the deposit, \ie{} in a vanishingly small fraction of the drop.  We expect negligible alteration of the stagnation flow.  Thus it is plausible that the altered height does not invalidate our predictions.  
			
\section{Conclusion} \label{sec:conclusion}

The fadeout mechanism described here represents a novel way to shape material in space using the generic properties of stagnation flow in an evaporating drop far from the site of deposition.  The mechanism is broadly applicable to a range of evaporation geometries to produce a controllable range of power law deposition profiles.  Further effects \eg from solute dispersion and packing will no doubt enrich the possibilities.  \revision{Such effects can be calculated explicitly and offer the prospect of further control.}  \omitt{5/26}

\acknowledgements

The author benefitted from discussions with Rui Zheng, Jin Wang, Wendy Zhang and David R. Nelson.  \revision{Adam Hammond and Vytas Bindokas  graciously provided the means } for making the micrograph of Fig. \ref{fig:concentrated}.   This work was supported in part by the National Science Foundation's MRSEC Program under Award Number DMR 0820054

\end{document}